\documentclass[conference]{IEEEtran}
\IEEEoverridecommandlockouts
\usepackage{cite}
\usepackage{amsmath,amssymb,amsfonts}
\usepackage{algorithmic}
\usepackage{graphicx}
\usepackage{textcomp}
\usepackage{xcolor}
\usepackage{booktabs}
\usepackage{multirow}
\usepackage{caption}
\usepackage{subcaption}
\usepackage[normalem]{ulem}
\useunder{\uline}{\ul}{}

\def\BibTeX{{\rm B\kern-.05em{\sc i\kern-.025em b}\kern-.08em
    T\kern-.1667em\lower.7ex\hbox{E}\kern-.125emX}}
\begin{document}

\title{Design Space Exploration for ReRAM-based Architectures to Address Scaling Non-idealities}
\author{
\IEEEauthorblockN{Ching-Yi Lin}
\IEEEauthorblockA{\textit{Department of Electrical and Computer Engineering} \\
\textit{University of Maryland}\\
College Park, USA
}
\and
\IEEEauthorblockN{Sahil Shah}
\IEEEauthorblockA{\textit{Department of Electrical and Computer Engineering} \\
\textit{University of Maryland}\\
College Park, USA
}
}
\maketitle
\begin{abstract}
ReRAM-based in-memory computing (IMC) architectures are promising candidates for energy-efficient matrix-vector multiplication. While scaling the size of ReRAM arrays allows for the amortization of power-hungry peripheral circuits like DACs and ADCs, it simultaneously introduces more parasitic along the signal path. Because of these challenges, current design methodologies often lack practical guidelines to balance these effects at early design stage, forcing designers to rely on time-consuming, iterative transistor-level simulations.

In this work, we propose a comprehensive framework for design space exploration that enables the selection of optimal array size, ADC resolution, and system frequency without requiring exhaustive simulations. The framework utilizes a specialized testbench to extract parameters from a limited set of representative transistor-level simulations. These parameters are then used to accurately predict the performance of arbitrary architectures. We demonstrate the effectiveness of this framework through two realistic design cases aimed at maximizing energy efficiency (TOPs/s/W). The results show that the framework successfully identifies optimal architectural configurations under strict power and error constraints, providing an efficient path for high-performance IMC design.
\end{abstract}


\section{Introduction}

\begin{figure}[b]
    \centering
    \includegraphics[width=\linewidth]{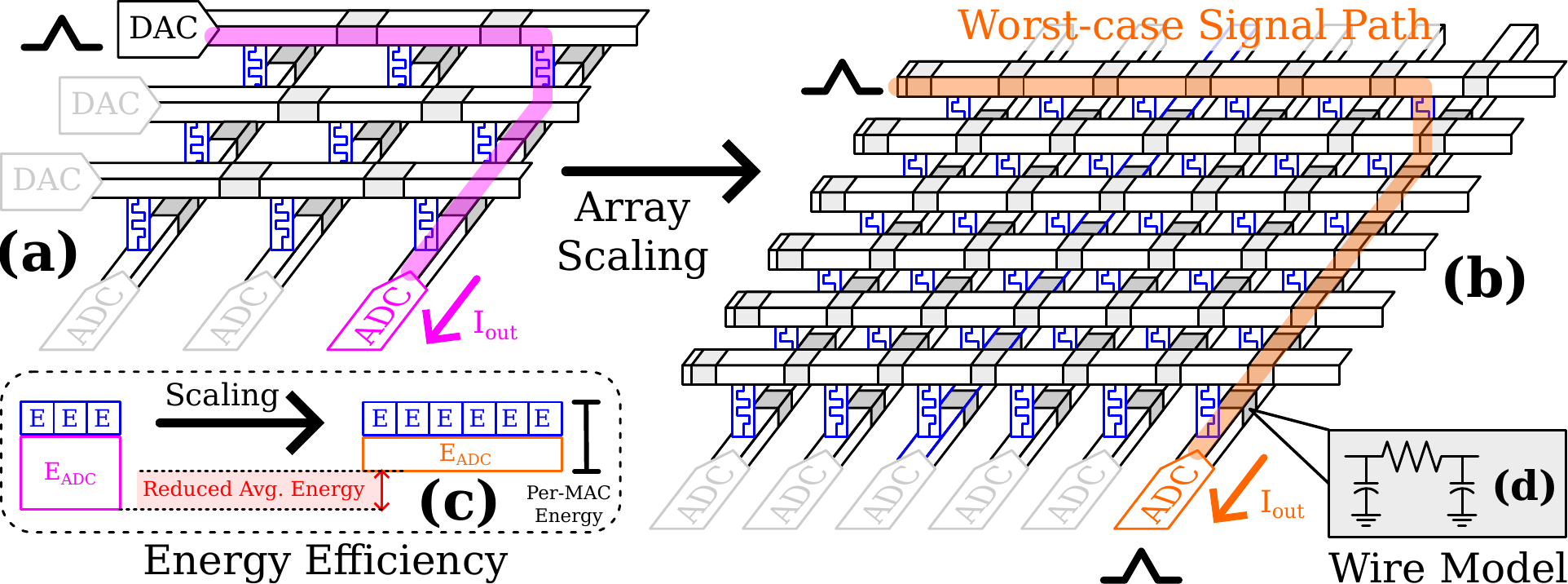}
    \caption{ReRAM array scaling on energy and parasitics: (a) A small $3\times 3$ array with 3 input DACs and 3 output ADCs. (b) Scaling to a $6\times 6$ array increases cell count by 4x but doubles ADCs and DACs. (c) Per-MAC energy improves as the fixed cost $E_{ADC}$ is amortized over a larger number of MACs (d) But double $\Pi$-model-based parasitic.}
    \label{fig:overview}
\end{figure}

Analog non-volatile memories are widely used to design in-memory computing architectures. Various devices have demonstrated the ability to both store data and perform computations \cite{baek_edge_2025}. Among these, ReRAM devices offer high area density and low read/write energy consumption \cite{didin_characterization_2026}. They allow weights to be programmed into different resistance states, such as the low-resistance state (LRS) and high-resistance state (HRS). A classic ReRAM-based in-memory computing structure is illustrated in Figure~\ref{fig:overview}(a) and Figure~\ref{fig:overview}(b): row-wise DACs provide input voltages, while column-wise ADCs collect the aggregated output current from ReRAM cells in the same column, following $I_j = \sum_i I_{ij}$. Since each ReRAM cell conducts current $I_{ij} = G_{ij} \times V_i$, this structure performs matrix multiplication $\mathbf{I} = \mathbf{G}\mathbf{V}$, enabling $N^2$ MAC operations to be executed in parallel.

Despite the high energy efficiency of ReRAM cells, ADCs and DACs introduce significant overhead in system-level energy consumption. For example, Yoon~\cite{yoon202140} presents an RRAM macro in which more than 50\% of the total power is consumed by peripheral circuits. One common approach to reducing this overhead is array scaling. In-memory computing arrays are typically evaluated in terms of per-MAC energy efficiency (e.g., TOPs/s/W or TOPs/J). Since the number of ADCs/DACs scales as $O(N)$, while the number of ReRAM cells scales as $O(N^2)$, increasing the array size $N$ amortizes peripheral costs and improves per-MAC energy efficiency. Figure~\ref{fig:overview}(c) illustrates this effect with a simple example: when an array is scaled from $3 \times 3$ to $6 \times 6$, even if the ADC energy $E_{\text{ADC}}$ remains constant, each ADC operation is amortized over more MAC operations, thereby reducing the energy per MAC at the system level.

Despite the energy-efficiency benefits of array scaling, the increased distance between sources (DACs) and destinations (ADCs) introduces new challenges. As shown in Figure~\ref{fig:overview}(b), the longest signal path in a double-sided array doubles when the array dimensions are doubled. These longer paths introduce parasitic resistances and capacitances, which not only slow down MAC operations but also reduce the effective voltage across ReRAM cells due to IR drop. These parasitic effects are typically modeled using a $\Pi$-shaped RC network, as illustrated in Figure~\ref{fig:overview}(d). According to the Elmore delay model~\cite{elmore1948transient}, doubling the array dimensions (width and height) can reduce the operating frequency by approximately $4\times$, despite the $4\times$ increase in the number of ReRAM cells.

Although several studies have explored array scaling and its negative impact on performance~\cite{lepri2022modeling,chen2023scaling}, few provide practical design guidelines for incorporating these effects during early-stage design. In particular, determining the optimal array size and operating frequency to maximize energy efficiency (TOPs/s/W) remains an open challenge. This gap between analysis and design forces designers to rely on iterative circuit simulations to evaluate non-idealities and select suitable architectural parameters. As a result, there is a need for a framework that enables designers to estimate system-level performance before detailed sub-module design begins.

In this work, we propose a framework to help designers select array size, ADC resolution, and system frequency without requiring extensive simulations. We first develop a testbench to characterize key parameters through transistor-level simulations of few representative architectures. These extracted parameters are then used for design space exploration to predict the performance of arbitrary architectures. To demonstrate the effectiveness of the proposed framework, we present two realistic design examples that maximize energy efficiency (TOPs/s/W) under power and error constraints.


\section{Background}
The effectiveness of ReRAM-based computing relies on its ohmic behavior, where the input voltage and output current maintain a constant conductance under ideal conditions. However, non-idealities in ReRAM arrays have been extensively studied from cell-level to system-level.

At the cell level, the I-V relationship is more precisely modeled as a $sinh$ function rather than a linear ohmic response~\cite{lentz2013current,messaris2018data}. Beyond deterministic behavior, statistical variations have also been widely investigated. Didin modeled resistance variations as a log-normal distribution \cite{didin_characterization_2026}, and Lin characterized ReRAM retention error as a mean shift with a constant rate in $g$-$log(t)$~\cite{lin2019performance}. As these statistical variations can be amortized in system-level analysis, this work focuses on the $sinh$ I-V relationship and adopts a Verilog-A implementation with nominal measurements.

At the system level, Roy first analyzed the effect of cell conductance variation and ADC/DAC noise in parasitic-free arrays~\cite{roy2022fundamental}. Lepri proposed a numerical algorithm to characterize IR drop effect in readout current and maximum error~\cite{lepri2022modeling}. Chen extended this analysis by incorporating IR drop and leakage current, modeling output flip probability as a function of array size and $R_{ReRAM}$~\cite{chen2023scaling}. In this work, we also develop a testbench to catch both IR drop and leakage current. Instead of binary outputs, we formulate error as an analog voltage difference. Additionaly, we provide a framework to enable designers to explore system-level performance tradeoffs in their early design stage.

\begin{figure}
    \centering
    \includegraphics[width=\linewidth]{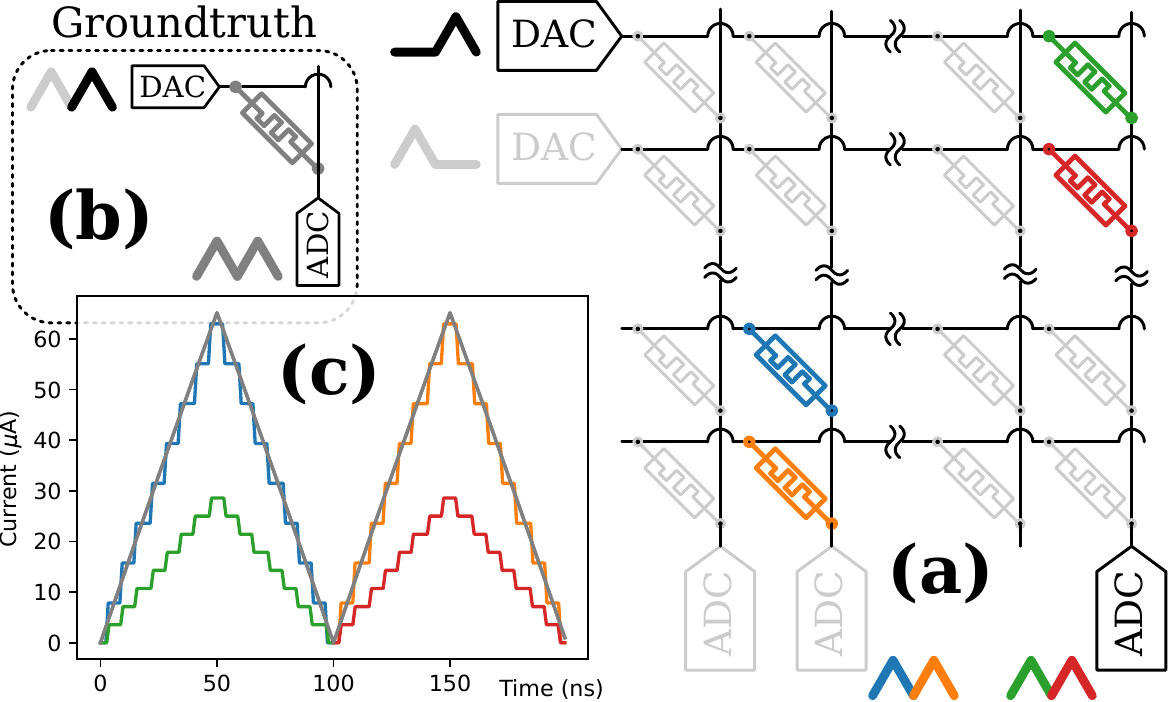}
    \caption{Testbench setup: (a) Each input DAC sequentially applies a triangular wave, and each output ADC collects the aggregated current (b) The reference current is obtained from an isolated, parasitic-free ReRAM cell with the same input (c) Transient current response for different locations within the array}
    \label{fig:testbench}
\end{figure}
\section{Testbench Setup}
\label{sec:testbench}
To evaluate per-cell performance within the array, the testbench is illustrated in Figure \ref{fig:testbench}(a). The array employs row-wise DACs to apply voltage input and column-wise ADCs to measure current output. Transient simulations are conducted over multiple time segments, where each segment activates a single row through a triangle wave while grounding the other rows. Corresponding ADCs then sample outputs at each column. With one-hot style input, each ADC output exhibits a repeating triangular wave in every time segments.

In this work, we focus on performance metrics in error and power. Error is defined as the root-mean-square current difference between the array outputs considering parasitics and a parasitic-free groundtruth, as shown in Figure \ref{fig:testbench}(b). Due to parasitics, cells located farther from DACs and ADCs experience more parasitics, resulting in increased errors. Figure \ref{fig:testbench}(c) demonstrates this phenomena, where far-end cells (red and green) exhibit greater deviation from the groundtruth (gray) compared to near-end cells (blue and orange).

\begin{figure}[tb]
    \centering
    \includegraphics[width=\linewidth]{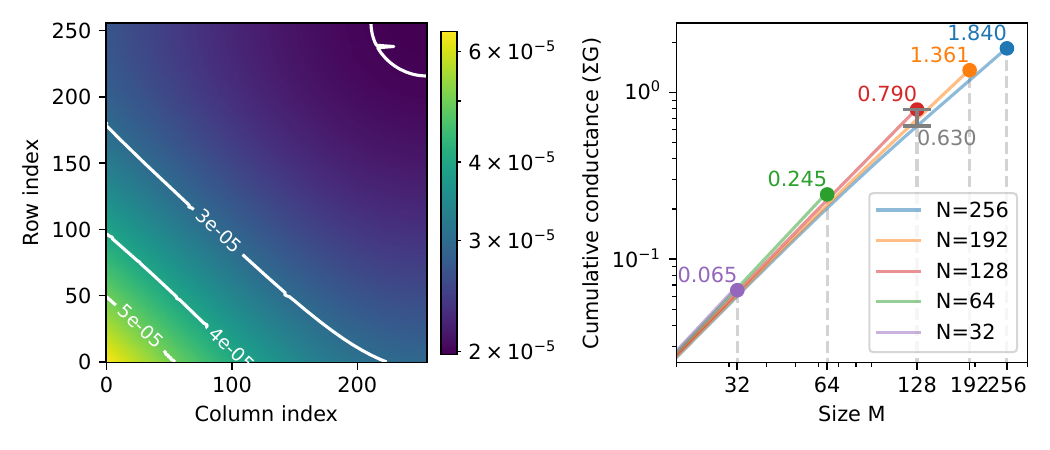}
    \caption{(a) Spatial distribution of $G_{eff}$ in a $256\times 256$ array (b) Cumulative conductance $\sum_{i}^{M}\sum_{j}^{M}G_{N,eff}[i,j]$ for varying N and M, demonstrating that the total conductance of an $M\times M$ array can be approximated using the extracted conductance $G_{N,eff}$ from a larger array.}
    \label{fig:geff}
\end{figure}

For power analysis, ReRAM power is estimated as the product of input voltage and output current. Although the actual power depends on both input voltage and ReRAM cell state, we adopt a worst-case assumption by setting $G_{ReRAM}=G_{LRS}$. Despite the fixed conductance in each ReRAM cell, effective conductance varies spatially due to its parasitic effects. We therefore extract per-cell effective conductance $G_{eff}=I_{OUT}/V_{IN}$ from the testbench for power estimation.
Figure \ref{fig:geff}(a) visualize a $256\times 256$ array of per-cell $G_{eff}$, where far-end cells exhibit lower conductance with the parasitic resistance involved in serial. Using per-cell conductance, total array power is estimated through
\begin{equation}
    P_{total} = \sum_{i,j} P_{i,j}=\sum_{i,j} (G_{eff}[i,j]\times{V_{IN}}^2)={V_{IN}}^2\times\sum_{i,j} G_{eff}[i,j]
    \label{eqn:pworst}
\end{equation}
This formulation separates architecture-dependent terms $\Sigma_{i,j} G_{eff}[i,j]$ from driver-dependent terms $V_{IN}^2$, enabling clearer analysis in subsequent sections.

\section{Simulation Result Generalization}
To generalize performance for an arbitrary $M\times M$ array based on simulattion of an $N\times N$ array in Section \ref{sec:testbench}, we analyze the scaling behavior of error and power in this section.

\begin{figure}[b]
    \centering
    \includegraphics[width=\linewidth]{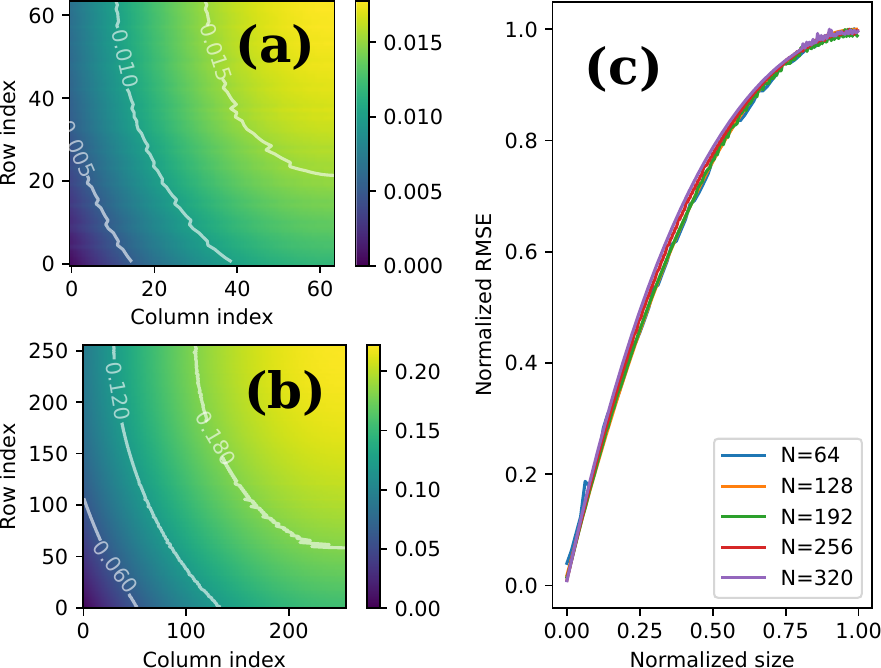}
    \caption{(a-b) RMSE distribution for a smaller $64\times 64$ array and a larger $256\times 256$ array. (c) Normalized RMSE as a function of normalized array size. The consistency of curves demonstrates the same error distribution for various array size $N$}
    \label{fig:rmse-normalized-allsize}
\end{figure}

\begin{figure}
    \centering
    \includegraphics[width=\linewidth]{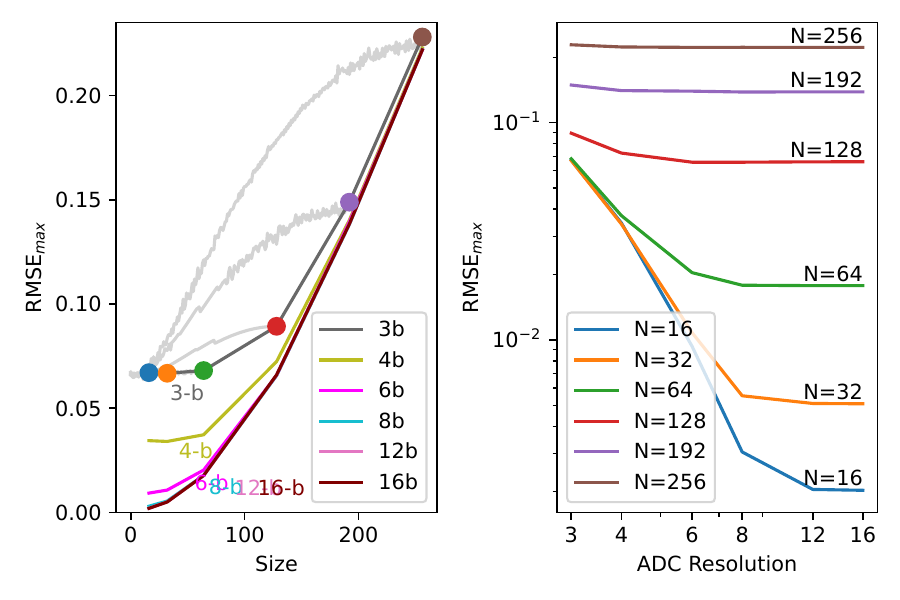}
    \caption{(a) RMSE-vs-size in different ADC resolution (b) RMSE-vs-resolution in different array size $N$}
    \label{fig:rmse-at0p2}
\end{figure}

The per-cell RMSE is visualized in 
Figure \ref{fig:rmse-normalized-allsize}(a) and Figure \ref{fig:rmse-normalized-allsize}(b) for small ($32\times 32$) and large ($256\times 256$) arrays. Despite difference in colorbar amplitude, both plots exhibit similar spatial patterns, suggesting a consistent error distribution. To verify this observation, we extract RMSE values along the diagonal axis and normalize both size and RMSE values by their individual maximum. The aligned curves in Figure \ref{fig:rmse-normalized-allsize}(c) indicates that normalized RMSE can be estimated for arbitrary square array sizes. To obtain the non-normalized RMSE, we estimate RMSE$_{max}$ to determine the scaling of the error distribution, Figure \ref{fig:rmse-at0p2}(a) visualizes the RMSE$_{max}$-size relationship. We observe the error in large arrays are more size-dominant since the main error source is the parasitic. In contrast, the errors in small arrays are more ADC-dominant. This can be rephrased as a more consistent error in large arrays in \ref{fig:rmse-at0p2}(b). 
In this work, we estimate the value RMSE$_{max}$ as a function of size and ADC resolution through interpolation. Notably, constructing an $N_{size}\times N_{resolution}$ grid requires only $N_{size}$ simulations, since ADC inputs can be reused across different ADC resolutions.



Power estimation is more straightforward given the $P_{total} = \Sigma_i P_i = V_{INPUT}^2\times\Sigma_i\Sigma_jG_{eff}[i,j]$. If we assume size-independent conductance $G_M[i,j] = G_N[i,j]$ in a $M\times M$ sub-array, the total power of this sub-array can be obtained through $P_{total,M}={V_{INPUT}}^2\times{\sum\limits_i}{\sum\limits_j}^{M\times M}G_{N,eff}[i,j]$ similar to Equation~\ref{eqn:pworst} but with different range. In this case, we only require $G_{N,eff}[\cdot,\cdot]$ from a large enough array to estimate power of any smaller arrays. However, this assumption slightly underestimates power due to increased parasitic effects in larger arrays. To be particular, Figure \ref{fig:geff}(b) plots the sub-array cumulative conductance $\Sigma_{i,j}G_{eff}[i,j]$ obtained from different $N\times N$ arrays. Here we can see summed conductance of a $128\times 128$ sub-array from a $256\times 256$ array yields 0.630S compared to 0.790S from direct simulation, resulting in 20.3\% difference in estimated power. To address this discrepancy, we build a lookup function for ${\sum\limits_{i,j}}^{M\times M}G_{M,eff}[i,j]$ through interpolation similar to RMSE. The values of this lookup function is equivalent to connecting each trace end at Figure \ref{fig:geff}(b).



\section{ReRAM Array Design Space Exploration}
In the previous sections, we generalized simulation results from few architectures to a broader design space. In this section, we present representative design examples to demonstrate how these results can be leveraged during the architectural design process.

In our framework, each ReRAM array architecture is characterized by three parameters: array size $N$, frequency $f$, and ADC resolution. Although the framework enables full design space exploration, we focus on two practical design examples to highlight its powerfulness and interpretability. While these examples focus on a power constraint and an error constraint separately, we emphasize that multiple constraints can be jointly applied in realistic scenarios.

\begin{figure}
    \centering
    \includegraphics[width=\linewidth]{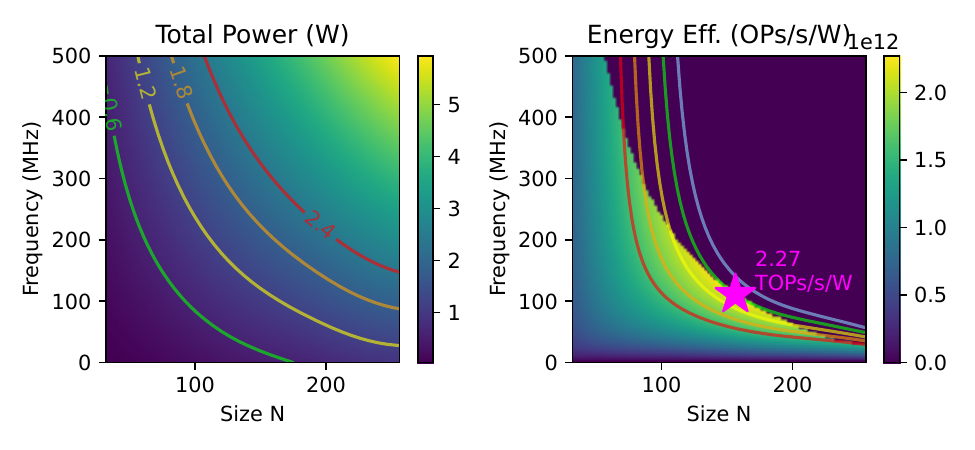}
    \caption{Size-frequency tradeoff under a power constraint: (a) Total power $P$ as a function of array size $N$ and frequency (b) Energy efficiency optimization under the $P<1.2$ W budget, achieving 2.27 TOPs/s/W with $N=156$ and $f=$111 MHz.}
    \label{fig:size-f-search}
\end{figure}

\subsection{Size-frequency Tradeoff with Power Constraint}

A common objective in architectural design is to identify the largest and fastest configuration under a given power budget. Although increasing both array size and operating frequency generally improves energy efficiency, determining the optimal combination under constraints is nontrivial. In this example, we aim to maximize energy efficiency subject to a power constraint of 1.2 W. Column readout employs 14-bit ADCs with an energy consumption of 39.19 pJ per conversion based on an ADC model~\cite{adc_survey}.

To determine the feasible design space, total power is estimated as
\begin{equation}
    P(N,f) = (\Sigma^{N\times N} G_{N,eff})\times V^2 + N\times E_{ADC}\times f
\end{equation}
With $\Sigma^{N\times N} G_i$ interpolated from Figure \ref{fig:geff}(b), the bivariate function can be visualized in Figure \ref{fig:size-f-search}(a). The power constraint corresponds to a boundary with negative slope.

To identify the optimal design under this power constraint, Figure \ref{fig:size-f-search}(b) calculates the energy efficiency within the feasible region. Along the constraint boundary, the iso-energy-efficiency contour forms a concave curve, with an optimal energy efficiency 2.27 TOPs/s/W achieved at an array size $N=156$ and $f=111$ MHz.

\subsection{Size-Resolution Tradeoff with Error Constraint}

Next, we consider a design scenario with an RMSE constraint of 0.2 when all ReRAM cells are set to the HRS, and a preset frequency of 300 MHz. For accuracy-sensitive application, higher ADC resolution is generally preferred. However, the exponential increase in ADC power with resolution poses a challenge for architectural optimization. While reducing array size can relax the required ADC resolution, its overall impact on energy efficiency is not immediately clear.

\begin{figure}[t]
    \centering
    \includegraphics[width=\linewidth]{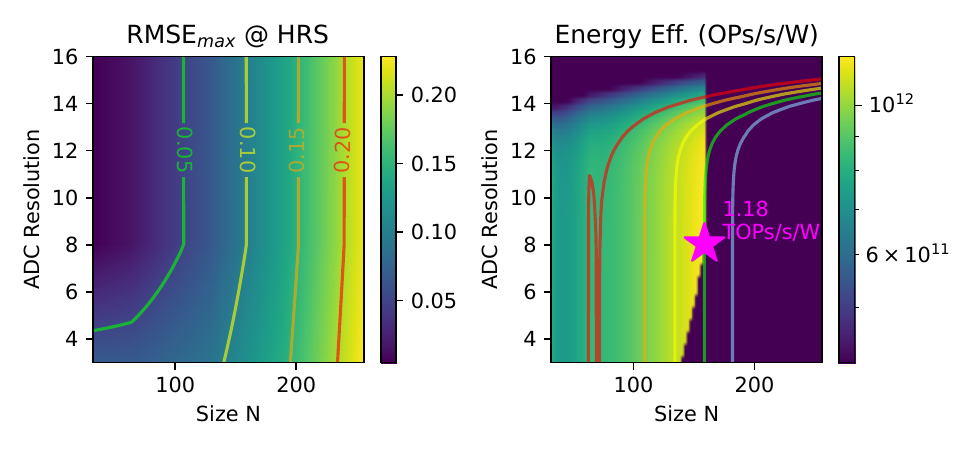}
    \caption{Size-resolution tradeoff under an error constraint: (a) Worst-case RMSE as a function of array size $N$ and ADC resolution (b) Energy efficiency optimization under the RMSE$_{max}< 0.1$ budget, achieving 1.18 TOPs/s/W at $N=159$ and 8-b ADC.}
    \label{fig:size-resol-search}
\end{figure}

Figure \ref{fig:size-resol-search}(a) visualizes RMSE as a function of array size and ADC resolution. The contour plot shows that, in larger arrays, the error is dominated by array size due to increased parasitics, as indicated by near-vertical contours. In contrast, for smaller arrays, ADC resolution plays a more significant role, since parasitic effects are less pronounced.

Following the same methodology as in the previous example, energy efficiency is evaluated within the feasible region defined by the RMSE constraint in Figure \ref{fig:size-resol-search}(b). Although the optimal design with $N=159$ and 8-b ADC lies on the constraint boundary, the energy efficiency landscape around this point is relatively flat. This behavior can be attributed to the reduced power dominance of lower-resolution ADCs and the resulting throughput-power cancellation in energy-efficiency calculation.

\section{Conclusions}
This paper presents a comprehensive framework for the design space exploration ReRAM-based in-memory computing architectures. While scaling ReRAM arrays improves energy efficiency by amortizing the power overhead of peripheral DACs and ADCs, it simultaneously introduces parasitic effects that can diminish these gains. To address this, our proposed framework enables to navigate these architectural tradeoffs from few simulation results in a specialized testbench. The effectiveness of the proposed framework was demonstrated through two realistic design examples: one optimizing size-frequency tradeoff under a power constraint, and another determining the ideal array size and ADC resolution to satisfy an error constraint. These results confirmed that our framework allows designers to identify optimal architecture and maximize the performance metrics during the early design stage.

\section{Acknowledgment}
This work was partly supported by both Semiconductor Research
Corporation (Award \#2023-AM-3160.032) and by the Army Research Office and was accomplished under Grant
Number W911NF-25-1-0260. The views and conclusions contained in this document are those of the authors and
should not be interpreted as representing the official policies, either expressed or implied, of the Army Research
Office or the U.S. Government. The U.S. Government is authorized to reproduce and distribute reprints for
Government purposes notwithstanding any copyright notation herein.

\bibliographystyle{IEEEtran}
\bibliography{references,references_shah}

\end{document}